\documentclass[onecolumn,11pt,aps,nofootinbib,superscriptaddress,tightenlines]{revtex4}
\usepackage{graphics,graphicx,psfrag,color,float, hyperref}
\usepackage{epsfig,bbold,bbm}
\usepackage{amsmath}
\usepackage{amssymb}

\usepackage{amsmath}
\usepackage{amssymb}
\usepackage{amsfonts}
\usepackage{amsthm, hyperref}
\usepackage {times}
\usepackage{amsfonts}

\newcommand{\beq}{\begin{equation}}
\newcommand{\enq}{\end{equation}}
\newcommand{\bel}{\begin{lemma}}
\newcommand{\enl}{\end{lemma}}
\newcommand{\bet}{\begin{theorem}}
\newcommand{\ent}{\end{theorem}}

\newcommand{\tr}{\mathrm{Tr}}
\newcommand{\myexp}{{\mathrm{e}}}

\newcommand{\eps}{\varepsilon}

\newcommand*{\cA}{\mathcal{A}}
\newcommand*{\cH}{\mathcal{H}}
\newcommand*{\cF}{\mathcal{F}}

\newcommand*{\cD}{\mathcal{D}}
\newcommand*{\cG}{\mathcal{G}}
\newcommand*{\cK}{\mathcal{K}}
\newcommand*{\cN}{\mathcal{N}}
\newcommand*{\cS}{\mathcal{S}}

\newcommand*{\cT}{\mathcal{T}}
\newcommand*{\cX}{\mathcal{X}}

\newcommand*{\cE}{\mathcal{E}}

\newcommand*{\cP}{\mathcal{P}}

\newcommand*{\cY}{\mathcal{Y}}
\newcommand*{\cR}{\mathcal{R}}

\newcommand{\bra}[1]{\langle #1|}
\newcommand{\ket}[1]{|#1 \rangle}
\newcommand{\braket}[2]{\langle #1|#2\rangle}

\newcommand*{\renyi}{R\'{e}nyi }

\newtheorem{theorem}{Theorem}
\newtheorem{lemma}{Lemma}

\begin {document}

\title{On the strong converses for the quantum channel capacity theorems}

\author{Naresh Sharma}
\email{nsharma@tifr.res.in}

\affiliation{School of Technology and Computer Science, \\
Tata Institute of Fundamental Research (TIFR), \\
Mumbai 400 005, India}

\author{Naqueeb Ahmad Warsi}
\email{naqueeb@tifr.res.in}

\affiliation{School of Technology and Computer Science, \\
Tata Institute of Fundamental Research (TIFR), \\
Mumbai 400 005, India}

\date{\today}

\begin{abstract}
A unified approach to prove the converses for the quantum channel capacity theorems is presented.
These converses include the strong converse theorems for classical or quantum
information transfer with error exponents
and novel explicit upper bounds on the fidelity measures reminiscent of the Wolfowitz
strong converse for the classical channel capacity theorems. We provide a new
proof for the error exponents for the classical information transfer.
A long standing problem in quantum information theory has been to find out the strong
converse for the channel capacity theorem when quantum information is sent
across the channel. We give the quantum error exponent thereby giving a one-shot
exponential upper bound on the fidelity. We then apply our results to show that the
strong converse holds for the quantum information transfer across an erasure channel
for maximally entangled channel inputs.
\end{abstract}

\maketitle

\section{Introduction}

One of the holy grails of information theory has been to prove the information-%
carrying capacities of various channels \cite{shannon1948}. The capacity identifies the
maximum rate (measured as number of bits/qubits per channel use) with which one
could transfer information reliably across the channel in the limit of sufficiently large
number of channel uses \cite{covertom,csiszar-korner-book,nielsen-chuang,hayashi,
wilde-book,gallager-68-book}.

The capacity, for certain channels,
also told us an interesting property about the fidelity  between the message
at the source and the replicated message at the receiver and this interesting property
is that the fidelity could be made (with appropriate inputs) to go to $1$
(i.e., a completely reliable transfer) for rates below capacity and goes to $0$
(i.e., a completely unreliable transfer)
for rates above capacity for any input for sufficiently large number of channel uses.

Such converse theorems where the fidelity goes to $0$ for sufficiently large number
of channel uses for rates above capacity are referred to as the strong converses.
For certain channels, one could show that the fidelity would decay exponentially to
zero as the number of channel uses increases for rates above capacity.  An example
of a channel with no strong converse is given in Ref. \cite{no-strong-converse-2011}.

A strong converse for the classical discrete memoryless
channel (DMC) was given by Wolfowitz \cite{wolfowitz-book}. In a simpler form, it showed
that for rates above capacity, $1-P_e$ ($P_e$ denotes the probability of decoding error)
can be bounded from above by two terms: one that
decays as $1/n$ and the other that decays exponentially
with $n$, where $n$ is the number of channel uses.

Arimoto provided a different strong converse with $P_e \to 1$ exponentially with $n$
using the error exponents \cite{arimoto-1973-converse}. These error
exponents were known from the work of Gallager who used them to give an upper 
bound to show $P_e \to 0$ exponentially for rates below capacity \cite{gallager-expo-1965}.

An important problem in quantum information theory has been to find the capacity
of a quantum channel for classical information transfer
\cite{holevo-hsw-1998,schu-west-hsw-1997,hastings-2009}. Winter provided
a strong converse which guarantees that for rates above capacity
$P_e \to 1$ as $n \to \infty$ \cite{winter-99-converse}.
Ogawa and Nagaoka gave an Arimoto-like strong converse
where they showed that $P_e \to 1$ exponentially with $n$ \cite{ogawa-1999-converse}.

The channel inputs used in the strong converse theorems mentioned above
were unentangled across channel uses. In a fully general channel converse, such a restriction
would not be made. K\"{o}nig and Wehner provide a strong converse for entangled inputs
for a subclass of channels for which a single-letter formula for capacity is available
\cite{konig-2009-converse}.
More strong converse theorems not necessarily in the context of channel
capacity could be found in Refs.
\cite{han-kobayashi-89-hypothesistesting, ogawa-nagaoka-99-converse, ogawa-nagaoka-2000-strong,
ahlswede-02-converse, ahlswede-mac-strong}.

Polyanskiy, Poor \& Verd\'{u} and Polyanskiy \& Verd\'{u} (see Refs. \cite{polyanskiy-2010-coding,
polyanskiy-2010-converse}) provided a unified converse
for the classical channel capacity theorem and such a converse yields among others the Arimoto converse
(Ref. \cite{arimoto-1973-converse}), Wolfowitz converse (Ref. \cite{wolfowitz-book})
and the Fano inequality (Ref. \cite{covertom}).

One of their building blocks has been the use of the monotonicity (or data processing 
inequality) of the divergences in the unified converse. It is interesting to note that
a similar approach was followed by Blahut in giving an alternate proof of the Fano inequality
in 1976 \cite{blahut1976}. This technique was also used
by Han and Verd\'{u} to generalize the Fano inequality \cite{han-verdu1994}.

Instead of relative entropy employed by Blahut, Polyanskiy-Poor-Verd\'{u}
used generalized divergences that satisfied the monotonicity and other properties.
The approach translated the promise of a communication protocol (or a code) of
delivering a rate under some fidelity constraints to an upper bound on the fidelity.
They also used some derived quantities defined by Csisz\'{a}r in
Ref. \cite{csizar-1995-rates},
who gave their operational characterizations in terms of block coding and
hypothesis testing and related it to the Gallager's error exponent. These quantities
play a critical role in the strong converse theorems.

We note that Csisz\'{a}r's approach in Ref. \cite{csizar-1995-rates} was generalized
to the quantum domain by Mosonyi and Hiai in Ref. \cite{mosonyi-2011}
to provide an operational interpretation of the quantum $\alpha$-relative entropies, but
there has been no connection made between the Csisz\'{a}r's quantities and the
strong converse theorems.

A related work has been the one done by the first author of using monotonicity for
proving the generalized quantum Fano inequality in Ref. \cite{sharmapra2008}.
Fano inequality is widely used in the converse (not strong) channel
capacity theorem proofs.
Note that classical Fano inequality has no special relation with the quantum Fano inequality
(the former not being a special case of latter) and the technique to generalize the quantum
Fano inequality is inherently `quantum' and perhaps the only common thread between the
quantum and the classical proof (the latter dating back to Blahut's work)
has been the use of monotonicity.

Our approach in this paper has been to carry this common thread of using monotonicity
further and to provide a unified approach to strong converses such as the quantum 
generalization of the Arimoto's and Wolfowitz's with explicit bounds for the latter. We note
that no quantum version of Wolfowitz-like bound was known. We
build on the above mentioned works in the
classical and quantum domains to first list the properties of generalized quantum divergences
that we shall need for our proofs. In particular, we show that the quantum \renyi divergences
and a non-commutative hockey-stick divergence, that we define, satisfy
these properties and suffice to give Arimoto-like
bounds with error exponents and also Wolfowitz-like bounds.

We then apply our approach to two quantum channel capacity theorems namely sending
classical information across a quantum channel and sending quantum information across
a quantum channel. We note that the strong converse for the latter problem has been an
open problem for quite sometime.

The organization of the paper is as follows. In Section \ref{div}, we list the properties
we desire from the generalized divergences that can be leveraged for the strong
converse theorems. We then derive quantities based on these divergences similar
to Csisz\'{a}r in Ref. \cite{csizar-1995-rates}.

In Section \ref{class-sec}, we first define the information-processing task for sending the classical information
across a quantum channel and then prove a converse for the generalized divergences. We then
take specific examples of divergences to yield the two converses - one coinciding with the
Ogawa and Nagaoka converse but with an alternate proof
and second which is Wolfowitz-like.

In Sec. \ref{quant-sec}, we repeat the above for sending the quantum information
across a quantum channel. The results are quantum error exponent reminiscent of
Gallager/Arimoto exponent and then Wolfowitz-like bounds. We give
sufficient conditions for the strong converse to hold in general. Lastly, we provide
a strong converse for the quantum erasure channel for maximally entangled channel inputs.

The proofs of many Lemmas are given in the Appendix to make the reading of the paper easier.
We use the following notation throughout the paper. All the logarithms are natural
logarithms. We shall assume that all the quantum systems are finite dimensional.
For a given Hilbert space $\cH_A$ describing quantum system $A$,
let $\cS(\cH_A)$ denote the set of all density matrices
of $\cH_A$ and let $|A|$ be the dimension of $\cH_A$.
${\mathbbm{1}}$ indicates the identity matrix whose dimensions would be
clear from the context. For a given square matrix $\rho$ and a scalar $x$, $\rho + x$ is
supposed to mean $\rho + x {\mathbbm{1}}$. A quantum operation is a completely
positive and trace preserving (CPTP) map and we use quantum operation, quantum
channel, and CPTP map synonymously.

The von Neumann entropy of a quantum state $\rho$ in system $A$ is denoted by $H(A)_\rho$ and if $\sigma^{AB}$
is a bipartite state in $AB$, then the quantum mutual information is given by
\beq
I(A;B)_\sigma := H(A)_\sigma + H(B)_\sigma - H(A,B)_\sigma.
\enq
The coherent information of $\sigma^{AB}$ is given by
\beq
I(A \rangle B)_\sigma := H(B)_\sigma - H(A,B)_\sigma.
\enq
The projector $P_{\{\rho-\sigma \geq 0\}}$ is a projector onto the positive part of $\rho-\sigma$.
For a pure state $\ket{\phi}$, we denote $\ket{\phi}\bra{\phi}$ by $\phi$.
The fidelity between a pure and a mixed state is defined as $F(\ket{\phi},\rho)$
$= \bra{\phi} \rho \ket{\phi}$.

\section{Generalized divergences}

\label{div}

Let  us denote a generalized divergence for positive matrices from
$\rho$ to $\sigma$ by $\cD(\rho || \sigma)$ that satisfies the following properties:
\begin{enumerate}
\item $\cD(\rho || \sigma)$ satisfies the monotonicity property (or the data processing inequality), i.e., for any CPTP map $\cE$, we have
\beq
\cD(\rho || \sigma) \geq \cD\left[\cE(\rho) || \cE(\sigma)\right].
\enq
\item For any quantum state $\kappa$,
\beq
\cD(\rho \otimes \kappa || \sigma \otimes \kappa) = \cD(\rho || \sigma).
\enq
\item Let $\Pi_0 = \ket{0}\bra{0}$ and $\Pi_1 = \ket{1}\bra{1}$ be two projectors with
$\Pi_0 + \Pi_1 = \mathbbm{1}$.

For $\alpha, \beta \in [0,1]$, let $\rho = (1-\alpha) \Pi_0 + \alpha \Pi_1$,
$\sigma = \beta \Pi_0 + (1-\beta) \Pi_1$, and let us define
\beq
{\mathbbm{d}}^{(c)}(1-\alpha || \beta) := \cD(\rho || \sigma).
\enq
Then ${\mathbbm{d}}^{(c)}(1-\alpha||\beta)$ is independent of the choice of
$\{\Pi_0,\Pi_1\}$ and decreasing for all $\alpha \leq 1 - \beta$.

Let $\alpha \in [0,1]$, $\beta \in (0,1]$, $\rho = \alpha \Pi_0 + (1-\alpha) \Pi_1$,
$\sigma = \beta \Pi_0 + (1/\beta-\beta) \Pi_1$, and let us define
\beq
{\mathbbm{d}}^{(q)}(\alpha || \beta) := \cD(\rho || \sigma).
\enq
Note that $\sigma \geq 0$ but does not have unit trace.
Then ${\mathbbm{d}}^{(q)}(\alpha||\beta)$ is independent of the choice of
$\{\Pi_0,\Pi_1\}$ and increasing for all $\alpha \geq \beta$.
\end{enumerate}
For our purposes, it is not necessary that all these properties are satisfied by a chosen divergence $\cD$.
Nevertheless, we give some examples below that satisfy all the above properties. \\

\noindent {\bf Some \renyi divergences}: For $\rho, \sigma \geq 0$, the \renyi divergence from $\rho$ to $\sigma$ of order $\alpha$,
$\alpha \in [0,2] \backslash \{1\}$,
is defined as
\beq
D_\alpha(\rho || \sigma) = {1 \over \alpha - 1} \log \tr \rho^{\alpha} \sigma^{1-\alpha}
\enq
and limit is taken at $\alpha = 1$. The monotonicity property is proved in Ref. \cite{hiai-2011} (see
Example 4.5) and the other properties are not difficult to show. \\

\noindent {\bf Non-commutative hockey-stick divergence}: It has been shown in Refs. \cite{polyanskiy-2010-coding,
polyanskiy-2010-converse} that the classical Wolfowitz converse giving explicit bounds
can be obtained using the $f$-divergence with
\beq
\label{hstick}
f(x) = (x-\gamma)^+,
\enq
where $x^+ = x$ if $x > 0$ and $0$ otherwise. This function is  known as the hockey-stick
function and has applications in finance \cite{hull-finance-book}.
It might be tempting to define a quantum
$f$-divergence (see Refs. \cite{ruskai-1999, sharma-qinf-2012, hiai-2011}) using $f(x)$ but $f(x)$ is not operator convex \cite{comment-op-convex}
and operator convexity is typically used for proving the monotonicity property.
However, there is a workaround. Let the Jordan decomposition of a square matrix $\kappa$ be given by
$\kappa = \kappa^+ - \kappa^-$,
where $\kappa^+$ and $\kappa^-$ are the positive and the negative parts of $\kappa$. Then
we could define the non-commutative hockey-stick divergence (or simply
hockey-stick divergence) as
\beq
\cD(\rho || \sigma) = \tr(\rho - \gamma \sigma)^+,
\enq
where $\gamma \geq 1$. Note that this is not a $f$-divergence in
the sense of \cite{ruskai-1999, sharma-qinf-2012, hiai-2011}. In fact, it is related to the trace distance since
$2(x-\gamma)^+ = |x - \gamma| + (x - \gamma)$, and hence, for quantum states $\rho$ and $\sigma$, we have
\begin{align}
2 \cD(\rho || \sigma) & = \tr| \rho - \gamma \sigma | + \tr( \rho - \gamma \sigma ) \\
& = || \rho - \gamma \sigma ||_1 + (1-\gamma).
\end{align}
The monotonicity follows from Lemma \ref{mono-hstick} in the Appendix and
the other properties are not difficult to show.

\subsection{Derived quantities}

We now define two quantities for bipartite states $\rho^{AB} \in \cS(\cH_A \otimes \cH_B)$
derived from the generalized divergence $\cD(\cdot||\cdot)$ as
\begin{align}
\cK^{(c)}(A;B)_\rho & := \inf_{\sigma^B \in \cS(\cH_B)} \cD(\rho^{AB} || \rho^A \otimes \sigma^B), \\
\cK^{(q)}(A;B)_\rho & := \inf_{\sigma^B \in \cS(\cH_B)} \cD(\rho^{AB} || \mathbbm{1} \otimes \sigma^B),
\end{align}
where $\rho^A = \tr_B \rho^{AB}$. We now have the following lemma that shows that
both the above derived quantities satisfy the data processing inequality.
\begin{lemma}
Let $\cE^{B \to C}$ be a CPTP map and $\rho^{AC} = \cE^{B \to C}(\rho^{AB})$. Then
\begin{align}
\label{temp1}
\cK^{(c)}(A;B)_\rho & \geq \cK^{(c)}(A;C)_\rho, \\
\label{temp2}
\cK^{(q)}(A;B)_\rho & \geq \cK^{(q)}(A;C)_\rho.
\end{align}
Let $\rho^{ABC} = \rho^{AB} \otimes \rho^C$. Then
\beq
\label{temp3}
\cK^{(c)}(A;B)_\rho \geq \cK^{(c)}(A;BC)_\rho.
\enq
\end{lemma}
\begin{proof}
For any $\delta > 0$, there exists a $\sigma^B$ such that $\cK^{(c)}(A;B)_\rho \geq \cD(\rho^{AB} || \rho^A \otimes \sigma^B) - \delta$.
We now have
\begin{align}
\cK^{(c)}(A;B)_\rho & \geq \cD(\rho^{AB} || \rho^A \otimes \sigma^B) - \delta \\
& \geq \cD \left[ \rho^{AC} || \rho^A \otimes \cE^{B \to C}(\sigma^B) \right] - \delta \\
& \geq \inf_{\sigma^C} \cD(\rho^{AC} || \rho^A \otimes \sigma^C) - \delta \\
& = \cK^{(c)}(A;C)_\rho - \delta.
\end{align}
Since this is true for any $\delta > 0$, the result follows. The proof of \eqref{temp2} is similar and we omit it.
To show \eqref{temp3}, note that
\begin{align}
\cK^{(c)}(A;B)_\rho & = \inf_{\sigma^B} \cD(\rho^{AB} || \rho^A \otimes \sigma^B) \\
& = \inf_{\sigma^B} \cD(\rho^{AB} \otimes \rho^C || \rho^A \otimes \sigma^B \otimes \rho^C) \\
& = \inf_{\sigma^B} \cD(\rho^{ABC} || \rho^A \otimes \sigma^B \otimes \rho^C) \\
& \geq \inf_{\sigma^{BC}} \cD(\rho^{ABC} || \rho^A \otimes \sigma^{BC}) \\
& = \cK^{(c)}(A;BC)_\rho.
\end{align}
QED.
\end{proof}
Note that the above definition of $\cK^{(c)}$ can be easily extended for the classical
random variables by assuming that the density matrices are commuting and the
random variables have probability distributions given by the eigenvalues.
This definition would be the same as the one in Ref. \cite{polyanskiy-2010-converse} given by
\beq
\label{class-k}
\cK^{(c)}(X;Y) = \inf_{Q_Y \in \cP_Y} \cD(P_{XY} || P_X \times Q_Y),
\enq
where $P_{XY}$ is the joint probability distribution of the pair of random variables
$(X,Y)$, $P_X$ is the probability distribution of $X$ that can be deduced from
$P_{XY}$, and $\cP_Y$ is the set of all probability distributions that $Y$ can take.
The following result will be useful later.
\begin{lemma}[Polyanskiy and Verd\'{u}, 2010 \cite{polyanskiy-2010-converse}]
\label{pverdu1}
Let the random variables $S, X, Y, \hat{S}$ form a Markov chain
$S - X - Y - \hat{S}$. Then
\beq
\cK^{(c)}(S;\hat{S}) \leq \cK^{(c)}(X;Y).
\enq
\end{lemma}

\section{Classical information over quantum channel}

\label{class-sec}

\subsection{Information processing task and converse}

For a given message source and a communication channel, a communication protocol
consists of an encoder and decoder entrusted with the task of replicating the message
at the receiver within some prescribed error.

Suppose Alice wants to send classical information to Bob using a quantum  channel.
We model the information as a  uniformly distributed random variable $S$ that takes values 
over the set $\{1, 2, ..., \myexp^{n\cR}\}$. Alice maps $S$ to $X$ using, possibly, a randomized 
encoder modeled by the conditional probability distribution $P_{X|S}$, where $X$ takes
values over $\{1,2,...,|\cX|\}$. The encoder's output is then mapped to
$\rho_X^{A^{\prime n}} \in \cS(\cH_{A^{\prime n}})$ and is sent to Bob over $n$ independent
uses of the channel $\cN^{A^\prime \to B}$. It is useful to represent the state
of input to the channel as a cq (classical-quantum) state given by
\beq
\rho^{MA^{\prime n}} = \sum_{x} P_{X}(x) \ket{x} \bra{x}^M \otimes \rho_x^{A^{\prime n}}.
\enq
Bob receives his part, $B^n$, of
\beq
\rho^{M B^n} = \cN^{A^{\prime n} \to B^n} (\rho^{MA^{\prime n}}),
\enq
where $\cN^{A^{\prime n} \to B^n} = \left(\cN^{A^\prime \to B} \right)^{\otimes n}$,
and to find out the classical message that Alice sent for him, Bob applies a
POVM $\{\Lambda_y^{B^n}\}$, $y \in \{1,2,...,|\cY|\}$, and the outcome of the
measurement process is modeled by a random variable $Y$ where
\beq
P_{Y|X}(y|x) = \Pr\{ Y = y | X = x\} = \tr \, \Lambda_y^{B^n}
\left(\cN^{A^\prime \to B} \right)^{\otimes n} \rho_x^{A^{\prime n}}.
\enq
The random variable $Y$ is then further processed (decoded) by Bob to yield $\hat{S}$ as an estimate
of the message $S$. The average probability of error is given by $\Pr\{S \neq \hat{S}\}$.

If the above communication protocol achieves an average probability of error not larger than
$\eps$, then we shall refer to such a protocol as a $(n,\cR,\eps)$ code.

We first prove an inequality similar to the Holevo bound for $\cK^{(c)}$.
\begin{lemma}[Holevo-like bound for $\cK^{(c)}$]
\label{gholevo}
For any $n$ and any POVM $\{\Lambda_y^{B^n}\}$, we have
\beq
\cK^{(c)}(X;Y) \leq \cK^{(c)}(M;B^n)_\rho.
\enq
\end{lemma}
\begin{proof}
We prove it for $n=1$ and the extension to any $n$ is straightforward.
Consider an ancilla quantum system $C$ that is uncorrelated with the system $MB$ and
the joint state of $MBC$ is given by
\beq
\rho^{MBC} = \sum_x P_X(x) \ket{x} \bra{x}^M \otimes \cN^{A^\prime \to B}
(\rho_X^{A^\prime}) \otimes \ket{1} \bra{1}^C,
\enq
where $\{\ket{i}^C\}$, $i=1,2,...,|C|$ is an orthonormal basis in $\cH_C$.
Let $\cE^{BC \to B'C'}$ be a CPTP map with
Krauss operators $\{\sqrt{\Lambda_y} \otimes U_y\}$, $y=1,2,...,|\cY|$, where $U_y$
is a Unitary matrix such that $U_y \ket{1}^C = \ket{y}^C$. The state after applying the map $\cE^{BC \to B'C'}$
is given by
\beq
\rho^{MB'C'} = \sum_{x,y} P_X(x) \ket{x} \bra{x}^M \otimes \sqrt{\Lambda_y}
\cN^{A^\prime \to B} (\rho_X^{A^\prime}) \sqrt{\Lambda_y} \otimes \ket{y} \bra{y}^C.
\enq
We now have the following inequalities
\begin{align}
\cK^{(c)}(M;B)_\rho & \stackrel{a}{\geq} \cK^{(c)}(M;BC)_\rho \\
& \geq \cK^{(c)}(M;B'C')_\rho \\
& \geq \cK^{(c)}(M ; C')_\rho,
\end{align}
where $a$ follows from \eqref{temp3}.
Note that
\beq
\rho^{MC'} = \sum_{x,y} P_X(x) P_{Y|X}(y|x) \, \ket{x} \bra{x}^M \otimes \ket{y} \bra{y}^C.
\enq
Let $\Pi_y = \ket{y} \bra{y}^C$ and let us define a quantum operation $\cF$ on the system
$C'$ with
Krauss operators $\{\Pi_y\}$. Since $\cF^{C' \to C''}(\rho^{MC'}) = \rho^{MC'}$, hence, using the
data processing inequality, for any $\sigma^{C'} \in \cS(\cH_{C'})$, we get
\beq
\cD(\rho^{MC'} || \rho^M \otimes \sigma^{C'}) \geq \cD\left[\rho^{MC'} || \rho^M \otimes
\cF^{C' \to C''}(\sigma^{C'}) \right].
\enq
This indicates that for the minimization, one may consider only those $\sigma^{C'}$ that
have $\{\ket{y}^C\}$ as the eigenvectors which would lead us to the classical divergence
in \eqref{class-k} and hence,
\beq
\cK^{(c)}(M ; C')_\rho = \cK^{(c)}(X ; Y).
\enq
QED.
\end{proof}

We now prove a theorem that would allow us to yield the various converses.
\begin{theorem}
\label{class-conv}
For $\eps \leq 1 - \myexp^{-n\cR}$, any $(n,\cR,\eps)$ code satisfies
\beq
{\mathbbm{d}}^{(c)}(1-\eps || \myexp^{-n\cR}) \leq \cK^{(c)}(M;B^n)_\rho.
\enq
\end{theorem}
\begin{proof}
We have the following inequalities
\begin{align}
\cK^{(c)}(M;B^n)_\rho & \stackrel{a}{\geq} \cK^{(c)}(X;Y) \\
& \stackrel{b}{\geq} \cK^{(c)}(S;\hat{S}) \\
& \stackrel{c}{\geq} {\mathbbm{d}}^{(c)}(\Pr\{S = \hat{S}\} || \myexp^{-n\cR}) \\
& \stackrel{d}{\geq} {\mathbbm{d}}^{(c)}(1 - \eps || \myexp^{-n\cR}),
\end{align}
where $a$ follows from Lemma \ref{gholevo}, $b$ follows from Lemma \ref{pverdu1},
$c$ follows by applying the classical transformation $(S,\hat{S}) \to \delta_{S,\hat{S}}$, where
$\delta_{x,y} = 1$ if $x=y$ and $0$ otherwise, and $d$ follows from Property 3 in Sec. \ref{div}
pertaining to $\mathbbm{d}^{(c)}$.
\end{proof}

It may be worth mentioning that the constraint  $\eps \leq 1 - \myexp^{-n\cR}$ may not be
seen as weakening the strong converse because, if
the constraint is violated, i.e., $\eps \geq 1 - \myexp^{-n\cR}$, then it, by itself,
would imply an exponential convergence of $\eps$ to $1$, where $\cR$ is bounded from below (since
we are proving the converse) by the channel capacity.

We are now in a position to apply Theorem \ref{class-conv} to yield the various
converses.

\subsection{New proof of the Ogawa and Nagaoka converse}

We assume that $n=1$ which is clearly the most general case.
Take $\cD$ to be $D_\lambda$, the \renyi divergence of order $\lambda \in $
$[0,2] \backslash \{1\}$, in Theorem \ref{class-conv} to get
\beq
\label{temp4}
\mathbbm{d}^{(c)}(1 - \eps || \myexp^{-\cR}) \leq \inf_{\sigma^B \in \cS(\cH_B)} D_\lambda(\rho^{MB} || \rho^M \otimes \sigma^B).
\enq
For a cq-state $\rho^{MB}$, we note from Ref. \cite{konig-2009-converse} that
\beq
D_\lambda(\rho^{MB} || \rho^M \otimes \sigma^B) = D_\lambda(\rho^{MB} || \rho^M \otimes 
\sigma^*) + D_\lambda(\sigma^* || \sigma^B),
\enq
where
\beq
\sigma^* = {\xi^B \over \tr \xi^B }, ~~~ \mbox{with} ~~~ \xi^B = \left( \sum_x P_X(x)
\rho^B_x \right)^\lambda.
\enq
Hence, the minimum of the RHS of \eqref{temp4} is achieved at $\sigma^B = \sigma^*$.
Substituting in \eqref{temp4}, we get
\beq
\mathbbm{d}^{(c)}(1 - \eps || \myexp^{-\cR}) \leq \frac{\lambda}{1-\lambda}
E_0(s,\cN^{A^\prime \to B})_{\{P_X(x),\rho_x^{A^\prime}\}},
\enq
where
\beq
E_0(s,\cN^{A^\prime \to B})_{\{P_X(x),\rho_x^{A^\prime}\}} =
-\log \tr \left\{ \sum_x P_X(x) \left[ \cN^{A^\prime \to B}(\rho_x^{A^\prime}) \right]^{1/(s+1)}
\right\}^{s+1}.
\enq
Using the inequality
\beq
\mathbbm{d}^{(c)}(1 - \eps || \myexp^{-\cR}) \geq {\lambda \over \lambda - 1}
\log(1 - \eps) + \cR,
\enq
we get for $\lambda \in (1,2]$, $s = \lambda^{-1} - 1$ and hence, $s \in [-1/2,0)$ that
\beq
\label{lowbnd1}
\eps > 1 - \exp\left\{-\left[ -s \cR + E_0(s,\cN^{A^\prime \to B})_{\{P_X(x),\rho_x^{A^\prime}\}}
\right] \right\},
\enq
The rest of the treatment is the same as in Ref. \cite{ogawa-1999-converse}.
For a $(n,\cR,\eps)$ code, let Alice send unentangled inputs across the channel uses,
i.e., the ensemble across the $n$ channel uses is given by
\beq
\label{unent-ens}
\left\{ \prod_{i=1}^n P_{X}(x_i),%
\bigotimes_{i=1}^n \cN^{A_i^\prime \to B_i}(\rho_{x_i}^{A_i^\prime}) \right\}, ~~~ x_i \in
\{1,2,...,|\cX|\}, ~~ \forall ~ i.
\enq

\begin{theorem}
For a $(n, \cR,\eps)$ code and for all $n$ $\in \mathbb{N}$
with unentangled inputs, the following lower bound holds
\beq
\label{temp9}
\eps \geq 1 - \exp \left\{-n\left[ -s \cR +
E_0(s,\cN^{A^\prime \to B})_{\{P_X(x),\rho_x^{A^\prime}\}} \right] \right\}.
\enq
\end{theorem}

It is also shown in Ref. \cite{ogawa-1999-converse} and not too difficult to check that
\begin{align}
\frac{ \partial E_0(s,\cN^{A^\prime \to B})_{\{P_X(x),\rho_x^{A^\prime}\}} } { \partial s}
\Big|_{s=0} & = I(M;B)_\rho,
\end{align}
and if $\cR > C^{(1)}(\cN)$ \cite{comment-additivity}, where
\beq
\label{temp10}
C^{(1)}(\cN) = \max_{\{P_X(x), \rho_x^{A^\prime}\}} I(M;B)_\rho,
\enq
then $\exists$ $t \in [-1/2,0)$ such that $\forall$ $s \in (-t,0)$,
\beq
-s \cR + E_0(s,\cN^{A^\prime \to B})_{\{P_X(x),\rho_x^{A^\prime}\}} > 0.
\enq
Hence, it implies using \eqref{temp9} that the probability of error
approaches $1$ exponentially.

Note that we had to confine $s$ in $[-1/2,0)$ instead of $[-1,0)$ since
the monotonicity of quantum \renyi divergence of order $\lambda$ is known to hold for
$\lambda \in [0,2]$ \cite{mosonyi-2011,hiai-2011}. This does not, however, affect the
strong converse proof since the Lemma 3 in Ref. \cite{ogawa-1999-converse}
would still hold.

\subsection{Wolfowitz converse}

\label{cwolf}

Again, we first assume $n=1$ before going to any $n$.
Take $\cD$ to be the hockey-stick divergence. We first note that
\begin{align}
\mathbbm{d}^{(c)}(1-\eps || \myexp^{-\cR}) & = \left(1 - \eps - \gamma \myexp^{-\cR}\right)^+ +
\left[\eps - \gamma (1-\myexp^{-\cR}) \right]^+ \\
\label{temp6}
& \geq 1 - \eps - \gamma \myexp^{-\cR}
\end{align}
and hence, using Theorem \ref{class-conv}, we have
\beq
\eps \geq 1 - \cK^{(c)}(M;B)_\rho - \gamma \myexp^{-\cR}.
\enq
Note that
\begin{align}
\cK^{(c)}(M;B)_\rho &
\leq \tr P_{\{\rho^{MB} - \gamma \rho^M \otimes \rho^B > 0\}} \rho^{MB}.
\end{align}
We now give an upper bound for the RHS of the above equation that is reminiscent of the
Chebyshev's inequality in the classical setting.
Using Lemma \ref{togetA}, we get for $\log\gamma > I(M;B)_\rho$,
\begin{align}
\label{temp7}
\tr P_{\{\rho^{MB} - \gamma \rho^M \otimes \rho^B > 0\}} \rho^{MB}
& \leq \frac{ \tr \rho^{MB} \left[ \log\rho^{MB}-\log(\rho^M \otimes \rho^B) \right]^2
- \left[ I(M;B)_\rho \right]^2 } {\left[\log\gamma - I(M;B)_\rho \right]^2},
\end{align}
Define for any $n \in \mathbb{N}$,
\beq
\cA^{(c)}_n := \max_{\rho^{MA^{\prime n}}} \left\{
\tr \rho^{MB^n} \left[ \log\rho^{MB^n}-\log(\rho^M \otimes \rho^{B^n}) \right]^2 -
\left[ I(M;B^n)_\rho \right]^2 \right\}.
\enq
This quantity (without the maximization over the channel input)
has been known in the classical case as the information variance and was defined by
Shannon (see Ref. \cite{polyanskiy-2010-coding} for more details).
The finiteness of $\cA^{(c)}_1$ follows from Lemma \ref{afinite}.
Using Theorem \ref{class-conv}, \eqref{temp6} and \eqref{temp7},
we get
\beq
\label{temp8}
\eps \geq 1 - \frac{\cA^{(c)}_1}{\left[\log\gamma - I(M;B)_\rho \right]^2} - \gamma
\myexp^{-\cR}.
\enq
For a $(n,\cR,\eps)$ code and unentangled inputs described in \eqref{unent-ens},
it is not difficult to show that $\cA^{(c)}_n = n \cA^{(c)}_1$. Then choosing
$\log\gamma = n C^{(1)}(\cN) + n \delta$ for some $\delta > 0$, where $C^{(1)}(\cN)$ is
defined in \eqref{temp10}, we get for this $(n,\cR,\eps)$
code
\beq
\eps \geq 1 - \frac{\cA^{(c)}_1}{n \delta^2} -  \myexp^{-n[\cR - C^{(1)}(\cN)-\delta]}.
\enq
Choosing $\delta = [\cR - C^{(1)}(\cN)]/2$, we get the following result.

\begin{theorem}
For a $(n,\cR,\eps)$ code with the ensemble given in \eqref{unent-ens},
the following lower bound holds
\beq
\label{temp11}
\eps \geq 1 - \frac{4\cA^{(c)}_1}{n [\cR - C^{(1)}(\cN)]^2} -  \myexp^{-n[\cR - C^{(1)}(\cN)]/2}.
\enq
\end{theorem}
Note that \eqref{temp11} has the same form as the classical
Wolfowitz strong converse (see Ref. \cite{gallager-68-book}).

\section{Quantum information over quantum channel}

\label{quant-sec}

\subsection{Information processing task and converse}

Suppose a quantum system $S$ and a reference system $A$ have a state $\ket{\phi}^{AS}$.
Alice only has access to the system $S$ and not to $A$. Alice wants to send her part of the
shared state with $A$ to Bob using $n$ independent uses of a quantum
channel $\cN^{A^{\prime} \to B}$ such that at the end of the communication protocol chain,
Bob's shared state with the reference $A$ is arbitrarily close to the state Alice shared
with $A$. We shall call $\cR$ to be the communication rate and is given by
\beq
\cR := \frac{ \log |S| }{n}.
\enq
We shall assume that the state of $S$ is given by
$\myexp^{- n\cR} {\mathbbm{1}}$, i.e., a completely mixed state.

To this end, Alice performs an encoding operation given by $\cE^{S \to A^{\prime n}}$ to get
\beq
\rho^{AA^{\prime n}} = \cE^{S \to A^{\prime n}} \left( \phi^{AS} \right).
\enq
Alice transmits the system $A^{\prime n}$ over
$\cN^{A^{\prime n} \to B^n} = \left(\cN^{A^\prime \to B} \right)^{\otimes n}$
and Bob receives the state
\beq
\rho^{AB^n} = \cN^{A^{\prime n} \to B^n} \left[ \cE^{S \to A^{\prime n}}
\left( \phi^{AS} \right) \right].
\enq
Bob applies a decoding operation on its part
of the received state to get
\beq
\rho^{A\hat{S}} = \cT^{B^n \to \hat{S}} \left\{
\cN^{A^{\prime n} \to B^n} \left[ \cE^{S \to A^{\prime n}} \left( \phi^{AS} \right) \right]
\right\}.
\enq
The performance of the protocol is quantified by the fidelity given by
\beq
F(\phi^{AS}, \rho^{A\hat{S}}) = \bra{\phi}^{AS} \rho^{A\hat{S}} \ket{\phi}^{AS}.
\enq
If we are given that the protocol achieves a fidelity not smaller than $\mathbb{F}$,
then we shall refer to such a protocol as a $(n,\cR,1-\mathbb{F})$ code.

The maximum rate per channel use for this protocol in the limit of large number of
channel uses and
fidelity arbitrarily close to $1$ was proved in a series of papers \cite{schumacher1996,
schumacher-nielsen-1996, barnum-nielsen-1998, barnum-knill-2000, lloyd-1997,
shor-cap-2002, devetak-2005, hayden-shor-2008}. Let the coherent information
of the channel $\cN^{A^\prime \to B}$ be defined as
\beq
Q(\cN) := \max_{\rho^{AA^\prime}} I(A \rangle B)_\sigma,
\enq
where $\sigma^{A B} = \cN^{A^\prime \to B} (\rho^{AA^\prime})$.
The capacity of the channel is now given by the regularization
\beq
Q_{\mathrm{reg}}(\cN) := \lim_{n \to \infty} \frac{ Q(\cN^{\otimes n}) } {n}.
\enq

We now prove a theorem that would give us one-shot inequalities between
the fidelity and the rate.
\begin{theorem}
\label{quant-conv}
For $\mathbb{F} \geq \myexp^{-n\cR}$, any $(n,\cR,1-\mathbb{F})$ code satisfies
\beq
\label{temp2000}
{\mathbbm{d}}^{(q)}(\mathbb{F} || \myexp^{-n\cR}) \leq \cK^{(q)}(A;B^n)_\rho.
\enq
\end{theorem}
\begin{proof}
Let $\{\ket{i}^{AS}\}$ be an orthonormal basis for $\cH_{AS}$ with
$\ket{1}^{AS} = \ket{\phi}^{AS}$. Consider a CPTP quantum map
$\cF^{A\hat{S} \to C}$ where $|C|=2$ with Krauss operators
$\ket{0}^C \bra{1}^{AS}$, and $\{\ket{1}^C \bra{i}^{AS} \}$, $i=2,3,...,|AS|$. Let
$\Pi_0^C = 0^C$ and $\Pi_1^C = 1^C$. Then we have
\begin{align}
\cF(\rho^{A\hat{S}}) & = \mathbb{F}^\prime \Pi^C_0 + (1-\mathbb{F}^\prime) \Pi^C_1, \\
\label{temp1001}
\cF({\mathbbm{1}} \otimes \sigma^{\hat{S}}) & = \myexp^{-n\cR} \Pi^C_0 +
(\myexp^{n\cR}-\myexp^{-n\cR})
\Pi^C_1,
\end{align}
where $\mathbb{F}^\prime = \bra{\phi}^{AS} \rho^{A\hat{S}} \ket{\phi}^{AS}$
and \eqref{temp1001} holds for all quantum states $\sigma^{\hat{S}}$.
We now have the following inequalities
\begin{align}
\cK^{(q)}(A;B^n)_\rho & \stackrel{a}{\geq} \cK^{(q)}(A;\hat{S})_\rho \\
& = \inf_{\sigma^{\hat{S}}} \cD(\rho^{A\hat{S}} || {\mathbbm{1}} \otimes \sigma^{\hat{S}}) \\
& \stackrel{b}{\geq} \inf_{\sigma^{\hat{S}}} \cD \left[ \mathbb{F}^\prime \Pi^C_0 + (1-\mathbb{F}^\prime) \Pi^C_1 || 
\myexp^{-n\cR} \Pi^C_0 + (\myexp^{n\cR}-\myexp^{-n\cR}) \Pi^C_1 \right] \\
& \stackrel{c}{=} {\mathbbm{d}}^{(q)}(\mathbb{F}^\prime || \myexp^{-n\cR}) \\
& \stackrel{d}{\geq} {\mathbbm{d}}^{(q)}(\mathbb{F} || \myexp^{-n\cR}),
\end{align}
where $a$ and $b$ follow from the data processing inequality, $c$ follows since the quantity
${\mathbbm{d}}^{(q)}(\mathbb{F}^\prime || \myexp^{-n\cR})$ is independent of $\sigma^{\hat{S}}$, and
$d$ from Property $3$ in Sec. \ref{div} pertaining to ${\mathbbm{d}}^{(q)}$.
\end{proof}

We now give upper bounds to the fidelity using the \renyi and the
hockey-stick divergences.

\subsection{Quantum Error exponent}

Let us first assume that $n=1$.
Take $\cD$ to be the \renyi divergence of order $\lambda \in [0,2] \backslash \{1\}$.
We first note that
\beq
\label{temp3002}
{\mathbbm{d}}^{(q)}(\mathbb{F} || \myexp^{-\cR}) \geq {\lambda \over \lambda - 1} \log \mathbb{F} + \cR
\enq
and it follows from Lemma \ref{qsibson} that
\beq
\cK^{(q)}(A;B)_\rho = {\lambda \over 1 - \lambda}
E_0(\lambda^{-1}-1, \cN^{A^\prime \to B})_\rho,
\enq
where for $s = \lambda^{-1} - 1$, $s \in [-1/2,0)$,
\beq
E_0(s, \cN^{A^\prime \to B})_\rho :=
-\log \tr \left\{ \tr_{A} \left[ \cN^{A^\prime \to B} (\rho^{AA^\prime}) \right]^{1/(1+s)}
\right\}^{s+1},
\enq
whose properties are studied by the following theorem.
\begin{theorem}
\label{qexp}
For any quantum state $\sigma^{AB}$, the function
\beq
g(s) := -\log \tr \left[ \tr_A \left( \sigma^{AB} \right)^{1/(1+s)} \right]^{s+1}, ~~ s \in [-1/2,0),
\enq
satisfies
\begin{align}
\label{zero}
g(0) & = 0, \\
\label{coherent}
\frac{\partial g(s)}{\partial s}\Big|_{s=0} & = I(A\rangle B)_{\sigma},
\end{align}
and $g(s) + (s+1) \log |A|$ is an increasing function in $s$.
\end{theorem}
\begin{proof}
See appendix.
\end{proof}

Using \eqref{temp2000}, we get the one-shot bound on the fidelity as
\beq
\label{temp2001}
\mathbb{F}\leq \exp\left\{-\left[-s\cR+E_0(s,\cN^{A^\prime \to B})_\rho \right] \right\}.
\enq
One could provide a sufficient condition for the
strong converse to exist as an additivity question. First define
\beq
E_0^*(s,\cN) := \min_{\rho^{AA^\prime}} E_0(s,\cN)_\rho,
\enq
where we just abbreviate $\cN$ for $\cN^{A^\prime \to B}$.
Then one could make the following statement. For $\cR \geq Q_{\mathrm{reg}}(\cN)$, strong
converse holds for all inputs if for all $m,n$ $\in \mathbb{N}$
$\exists$ $t \in [-\frac{1}{2},0)$ such that $\forall$
$s \in (t,0)$, 
\beq
\label{temp2002}
E_0^*\left(s,\cN^{\otimes n+m} \right) =  E_0^*\left(s,\cN^{\otimes n} \right) +
E_0^*\left(s,\cN^{\otimes m} \right).  
\enq
This statement is easy to prove. Using \eqref{temp2001}, we have
\beq
\mathbb{F} \leq \exp\left\{-n\left[-s\cR + \frac{E_0^*(s,\cN^{\otimes n})_\rho}{n} \right] \right\}.
\enq
If \eqref{temp2002} is satisfied, then $\forall$ $s \in (t,0)$, 
\beq
\frac{E_0^*(s,\cN^{\otimes n})_\rho}{n} = E_0^*(s,\cN)_\rho.
\enq
It follows from Lemma \ref{coh-lemma} that $\exists$ $t^\prime \in [-\frac{1}{2},0)$ such that
$\forall$ $s\in(t^\prime,0)$, $-s \cR + E_0^*(s,\cN)_\rho > 0$. Hence $\forall$
$s\in (\max\{t,t^\prime\},0)$ and $\forall$ $n$,
\beq
-s\cR + \frac{E_0^*(s,\cN^{\otimes n})_\rho}{n} >0
\enq  
and is independent of $n$. \eqref{temp2002} is unlikely to hold in general. Observe
that if the dimension of the quantum system $A$ is not constraining, then \eqref{temp2002}
implies the additivity of the coherent information of the channel.
To see this, divide \eqref{temp2002}
by $s$ and take the limit $s \uparrow 0$, invoke Theorem \ref{qexp}
and since $s < 0$, the minimum would be replaced by maximum over the input
states. It would be interesting to find out if there is a class of channels for which
\eqref{temp2002} holds.

\subsection{Wolfowitz converse}

Take $\cD$ to be the hockey-stick divergence.
Following the same steps as in \ref{cwolf}, we get for $n=1$
\begin{align}
\label{temp3003}
\mathbbm{d}^{(q)}(\mathbb{F} || \myexp^{-\cR}) & \geq \mathbb{F} - \gamma \myexp^{-\cR}, \\
\cK^{(q)}(A;B)_\rho & \leq \tr  P_{\{\rho^{AB} - \gamma \mathbbm{1} \otimes \rho^B > 0\}} \rho^{AB}.
\end{align}
We  upper bound the RHS of the above equation
using Lemma \ref{togetA}, for $\log\gamma > I(A \rangle B)_\rho$ as
\begin{align}
\label{temp12}
\tr P_{\{\rho^{AB} - \gamma \mathbbm{1} \otimes \rho^B > 0\}} \rho^{AB} & \leq
\frac{ \tr \rho^{AB} \left[ \log\rho^{AB}-\log(\mathbbm{1} \otimes \rho^B) \right]^2 -
\left[ I(A \rangle B)_\rho \right]^2 } {\left[\log\gamma - I(A \rangle B)_\rho \right]^2},
\end{align}
we get
\beq
\label{temp13}
\mathbb{F} \leq \frac{\cA_1^{(q)}}{\left[\log\gamma - I(A \rangle B)_\rho \right]^2} + \gamma \myexp^{-\cR},
\enq
where
\beq
\cA_n^{(q)} = \max_{\rho^{AA^{\prime n}}} \left\{ \tr \rho^{AB^n} \left[ \log\rho^{AB^n}
-\log(\mathbbm{1} \otimes \rho^{B^n}) \right]^2 - \left[ I(A \rangle B^n)_\rho \right]^2 \right\}.
\enq
For a $(n,\cR,1-\mathbb{F})$ code, choosing
$\log\gamma = n Q_{\mathrm{reg}}(\cN) + n \delta$ for some $\delta > 0$, we get
an upper bound
\beq
\mathbb{F} \leq \frac{\cA_n^{(q)}}{n^2 \delta^2} +
\exp \left\{-n[\cR - Q_{\mathrm{reg}}(\cN)-\delta] \right\}.
\enq
Choosing $\delta = [\cR - Q_{\mathrm{reg}}(\cN)]/2$, we get
\beq
\label{temp14}
\mathbb{F} \leq \frac{4\cA_n^{(q)}}{n^2 [\cR - Q_{\mathrm{reg}}(\cN)]^2} +
\exp \left\{- \frac{n}{2} [\cR - Q_{\mathrm{reg}}(\cN)] \right\}.
\enq
Hence, a sufficient condition for the strong converse to hold is that 
$\cA_n^{(q)}/n^2 \rightarrow 0$ as $n \rightarrow \infty$.

\subsection{Strong converse for the quantum erasure channel for maximally
entangled channel inputs}

A quantum erasure channel $\cN_{p}^{A^\prime \to B}$, defined in
Ref. \cite{grassl-1997}, is given by the following Krauss operators
$\left\{ \sqrt{(1-p)} \ket{i}^B\bra{i}^{A^\prime}, \sqrt{p}\ket {e}^B\bra{i}^{A^\prime}\right\}$, 
$i = 1,..., |A^\prime|$, $p \in [0,1]$, $|B| = |A^\prime|+1$, 
$\left\{\ket{i}^{A^\prime}\right\}, \left\{\ket{i}^B\right\}$ are orthonormal basis in
$\cH_{A^\prime}$ and $\cH_B$ respectively, and $\ket{e}^B = \ket{j}^B$ for
$j = |B|$. The action of the channel  can be understood as follows 
\beq
\cN_{p}^{{A^\prime} \to B}(\rho^{A{A^\prime}}) = (1-p) \sigma^{AB} + p \rho^A \otimes \ket{e} \bra{e}^B.
\enq 
Let $\sigma^{AB}$ $= \cG^{A^\prime \to B} (\rho^{AA^\prime})$, where $\cG$
increases the dimension but leaves the state intact.
Then with probability $1-p$, the channel 
leaves the state as $\sigma^{AB}$
and with probability $p$ it erases the state and replaces by $\ket{e}^B$.
It is not difficult to see that $\sigma^{AB}$ is orthogonal
to $\rho^A \otimes \ket{e} \bra{e}^B$.

One could carry over this observation for $2$ channel uses. Let
$\sigma^{AB^2}$ $= ( \cG^{A^\prime \to B} )^{\otimes 2} (\rho^{AA^{\prime 2}})$.
Observe that
\begin{align}
\left(\cN_{p}^{{A^\prime}_1 \to B_1} \otimes \cN_{p}^{{A^\prime}_2 \to B_2}\right)
(\rho^{A{A^\prime}_1{A^\prime}_2}) &= 
(1-p)^2 \sigma^{AB_1B_2} + p(1-p) \sigma^{AB_1} \otimes \ket{e} \bra{e}^{B_2}
\\ \nonumber
&\hspace{4mm} + p(1-p) \sigma^{AB_2} \otimes \ket{e} \bra{e}^{B_1} +
p^2 \rho^A  \otimes \ket{e} \bra{e}^{B_1} \otimes \ket{e} \bra{e}^{B_2},
\end{align}
where we use the usual notation for the reduced density matrices,
i.e., for example, representing
$\sigma^{A B_1}$ as the result of partial trace over $B_2$ of $\sigma^{AB^2}$,
each of these four matrices are orthogonal to each other and we have an abuse of
notation in the third term by rearranging the order of the systems.

Taking this further for $n$ channel uses, let $\sigma^{AB^n}$
$= ( \cG^{A^\prime \to B} )^{\otimes n} (\rho^{AA^{\prime n}})$.
The output can be written as the sum of $2^n$ orthogonal density
matrices where each of these matrices results from $i$ erasures $i \in\left\{0,...,n \right\}$
and this occurs with probability $(1-p)^{n-i}p^i$. The number of states that have
suffered exactly $i$ erasures is ${n \choose i}$.

Let  $B_{i_1} B_{i_2} \cdots B_{i_{n-k}}$ be the
quantum systems that have not suffered erasures and we could write the state in this
case using $\sigma^{AB^n}$ as
\beq
\zeta_{i_1,...,i_{n-k}}^{AB_{i_1} B_{i_2} \cdots B_{i_n}}
= \sigma^{AB_{i_1} B_{i_2} \cdots B_{i_{n-k}}} \otimes
\bigotimes_{j=1}^k \ket{e} \bra{e}^{B_{i_{n-k+j}}}.
\enq
It now follows that
\beq
\rho^{A B^n} = \sum_{2^n {\mathrm{terms}}}
\alpha_{k,n}  \times \zeta_{i_1,...,i_{n-k}}^{AB_{i_1} B_{i_2} \cdots B_{i_n}},
\enq
where
\beq
\alpha_{k,n} = (1-p)^{n-k} p^k.
\enq
To prove the strong converse, we find an upper bound for $\cK^{(q)}(A;B^n)$.
We assume that $\rho^{A A^{\prime n}}$ is a maximally entangled state with a Schmidt
rank of $|A|$.
Let the channel input be $d_A$ dimensional and
for $n$ channel uses, $|A| = d_A^n$. It is known that
$Q(\cN) = (1-2p)^+ \log d_A$ is the single-letter
quantum capacity for this channel \cite{bennett-1997} (see
also Ref. \cite{wilde-book}).

Note that
$d_A^k \times \rho^{A A^\prime_1 \cdots A^\prime_{n-k}}$ is a projector of
rank $d_A^k$. This may be a well-known observation and
is not difficult to prove but
for the sake of completeness, we provide a proof in the appendix in Lemma \ref{maxent}.
Observe that $\rho^{A^\prime_1 \cdots A^\prime_{n-k}}$ is
the maximally mixed state.
We note that the capacity-achieving input is maximally entangled.

For reasons that should be apparent from what follows, we also consider
$\lambda$-quasi relative entropy for $\lambda$ $\in [0,2]$ $\backslash \{1\}$ given by
\beq
\label{quasi}
D^{\mathrm{quasi}}_\lambda(\rho || \sigma) :=
\mathrm{sign}(\lambda-1) \tr \rho^\lambda \sigma^{1-\lambda}.
\enq
This relative entropy is jointly convex in its arguments and satisfies monotonicty
for the chosen range of $\lambda$ \cite{mosonyi-2011}. The \renyi divergence is a function
of this quantity.

For the hockey-stick divergence and $\lambda$-quasi relative entropy, note the
following identical steps
\begin{align}
\cD( \rho^{A B^n} || \mathbbm{1} \otimes \rho^{B^n} )
& \stackrel{a}{=} \sum_{2^n {\mathrm{terms}}}
\alpha_{k,n} \cD ( \zeta_{i_1,...,i_{n-k}}^{AB_{i_1} \cdots B_{i_n}} ||
\mathbbm{1} \otimes \zeta_{i_1,...,i_{n-k}}^{B_{i_1} \cdots B_{i_n}} ) \\
&  \stackrel{b}{=} \sum_{2^n {\mathrm{terms}}}
\alpha_{k,n} \cD ( \sigma^{AB_{i_1} \cdots B_{i_{n-k}}} ||
\mathbbm{1} \otimes \sigma^{B_{i_1} \cdots B_{i_{n-k}}} ) \\
\label{temp3000}
&  \stackrel{c}{\leq} \sum_{2^n {\mathrm{terms}}}
\alpha_{k,n} \cD ( \rho^{AA^\prime_{i_1} \cdots A^\prime_{i_{n-k}}} ||
\mathbbm{1} \otimes \rho^{A^\prime_{i_1} \cdots A^\prime_{i_{n-k}}} ),
\end{align}
where $a$ follows from orthogonality of $\zeta$'s, $b$ follows since we have
removed the tensors with $\ket{e}\bra{e}$, and $c$ follows from monotonicity.

The quantity $\cK^{(q)}(A;B^n)$ can be upper bounded by
$\cD( \rho^{A B^n} || \mathbbm{1} \otimes \rho^{B^n} )$.
For the \renyi divergence of order $\lambda \in (1,2]$, $\cK^{(q)}(A;B^n)$
is upper bounded by first computing \eqref{temp3000}
for the $\lambda$-quasi relative entropy to get
\beq
\cK^{(q)}(A;B^n) \leq \frac{n}{\lambda-1} \log
\left[ p d_A^{1-\lambda} + (1-p) d_A^{\lambda-1} \right].
\enq
Using \eqref{temp3002}, we get
\beq
\mathbb{F} \leq \exp \left( -\frac{\lambda-1}{\lambda} n
\left\{ \cR - \frac{\log\left[ p d_A^{1-\lambda} + (1-p) d_A^{\lambda-1} \right]}
{\lambda-1} \right\} \right).
\enq
The function
\beq
h(x) := \log\left[ p d_A^{1-x} + (1-p) d_A^{x-1} \right]
\enq
satisfies $h(1) = 0$. Furthermore, for $p \in [0,1/2]$,
\beq
\lim_{\lambda \downarrow 1} \frac{h(\lambda)}{\lambda-1} = Q(\cN).
\enq
Hence, for all $\cR > Q(\cN)$, $\exists$ $\lambda \in (1,2]$ s.t.
$\cR - h(\lambda)/(\lambda-1)$ $> 0$, and thus the strong converse holds.
For $p > 1/2$, $h^\prime(1)$ $< 0$ and hence, using similar arguments as above,
the strong converse follows.

For the hockey-stick divergence, \eqref{temp3000} is computed as
\begin{align}
\cK^{(q)}(A;B^n)
& \leq \sum_{k=0}^n
{n \choose k} \alpha_{k,n} \tr \Big( \rho^{AA^\prime_1 \cdots A^\prime_{n-k}} -
\gamma \mathbbm{1} \otimes \rho^{A^\prime_1 \cdots A^\prime_{n-k}} \Big)^+ \\
& \leq \sum_{k=0}^{\frac{n}{2} - \lfloor \frac{\log \gamma}{2\log d_A} \rfloor }
{n \choose k} \alpha_{k,n},
\end{align}
where we have upper bounded $\tr ( \rho^{AA^\prime_1 \cdots A^\prime_{n-k}} -
\gamma \mathbbm{1} \otimes \rho^{A^\prime_1 \cdots A^\prime_{n-k}} )^+$ by $1$
for $k \leq n/2 - \lfloor \log \gamma/(2\log d_A) \rfloor$.

Let us choose $\log \gamma = n[\cR + Q(\cN)]/2$ in \eqref{temp3000}.
For $\cR > Q(\cN)$, we have $n/2 - \lfloor \log \gamma/(2\log d_A) \rfloor < np$.
Using the Chernoff bound and \eqref{temp3003}, we get
\beq
\mathbb{F} \leq \exp \left\{-\frac{n}{2}[\cR-Q(\cN)] \right\} +
\exp \left\{ -\frac{n}{2p} \left[ \frac{(2p-1)^+}{2} + \frac{\cR}{4 \log d_A} \right]^2 \right\},
\enq
which gives us the strong converse.

\section{Acknowledgement}

The authors gratefully acknowledge the comments by A. Winter.


\begin{thebibliography}{10}
\bibitem{comment-op-convex}

\newblock A function $f(x)$ is said to be operator convex if for any two
  Hermitian matrices $\rho$ and $\sigma$ with spectrum in the domain of $f$ and
  any $\lambda \in [0,1]$, we have $f \left[ \lambda \rho + (1-\lambda) \sigma
  \right]$ $\leq$ $\lambda f(\rho)$ $+ (1-\lambda) f(\sigma)$, where we say
  $\rho \leq \sigma$, if $\sigma - \rho$ is positive semi-definite
  \cite{bhatia-1997}.
\end{thebibliography}

\appendix

\section{Proof of Lemmas}

\begin{lemma}
\label{mono-hstick}
Consider the matrices $\rho, \sigma \geq 0$ and a scalar $\gamma > 0$. Then for any CPTP map
$\cE$,
\beq
\tr(\rho - \gamma \sigma)^+ \geq \tr \left[ \cE(\rho) - \gamma \cE(\sigma) \right]^+.
\enq
\end{lemma}
\begin{proof}
Let the Jordan decomposition of $\rho - \gamma \sigma = Q - S$, where $Q,S \geq 0$. Let
$P := P_{\{ \cE(\rho) - \gamma \cE(\sigma) \geq 0 \}}$. Then
\begin{align}
\tr(\rho - \gamma \sigma)^+ & = \tr Q \\
& \stackrel{a}{=} \tr \cE(Q)  \\
& \stackrel{b}{\geq} \tr P [\cE(Q) - \cE(S)]  \\
& = \tr \left[ \cE(\rho) - \gamma \cE(\sigma) \right]^+,
\end{align}
where $a$ follows since $\cE$ is trace preserving, $b$ follows since
we are subtracting non-negative terms.
\end{proof}

\begin{lemma}
\label{positivity}
Let $f: \mathbb{R}^+ \to \mathbb{R}$ be an operator monotone function.
Then for $\rho, \sigma \geq 0$, $P := P_{\{\rho-\sigma\geq 0\}}$, we have
\beq
\tr P\rho\left[f(\rho)-f(\sigma)\right] \geq 0.
\enq
\end{lemma}
\begin{proof}
We follow the arguments similar to Theorem 11.18 in Ref. \cite{petz-book}
(see also the proof of Lemma 1 in Ref. \cite{kmr-2006-chernoff}).
Using the L\"{o}wner's Theorem (see Ref. \cite{petz-book}),
\beq
f(x) = \int_0^\infty \frac{x(1+\lambda)}{x+\lambda} d \mu(\lambda),
\enq
where $\mu$ is a positive finite measure, we get
\beq
\tr P\rho\left[f(\rho)-f(\sigma)\right] = \int_0^\infty
\lambda (1+\lambda) \tr\left[ P \rho (\rho+\lambda)^{-1} (\rho - \sigma)
(\sigma + \lambda)^{-1} \right] d\mu(\lambda).
\enq
To prove the inequality, it is sufficient to show that
$ \tr P \rho (\rho+\lambda)^{-1} (\rho - \sigma) (\sigma + \lambda)^{-1} \geq 0$
$\forall$ $\lambda > 0$. For $\Delta = \rho - \sigma$, we can get the integral representation
\beq
\tr P \rho (\rho+\lambda)^{-1} (\rho - \sigma) (\sigma + \lambda)^{-1}
= \int_0^1 \lambda \tr\left[ P \rho (\sigma+t \Delta+\lambda)^{-1} \Delta
(\sigma + t \Delta +\lambda)^{-1} \right] dt.
\enq
Hence, it is sufficient to show that
\beq
\label{temp15}
\tr P \rho (\sigma+t \Delta+\lambda)^{-1} \Delta (\sigma + t \Delta +\lambda)^{-1} \geq 0.
\enq
It is shown in Theorem 11.18 in Ref. \cite{petz-book} that
$\tr P \sigma (\sigma+t \Delta+\lambda)^{-1} \Delta
(\sigma + t \Delta +\lambda)^{-1} \geq 0$. Now it is easy to see that
$\tr P \Delta (\sigma+t \Delta+\lambda)^{-1} \Delta
(\sigma + t \Delta +\lambda)^{-1}$ $=\tr P 
\left[ \Delta (\sigma + t \Delta +\lambda)^{-1} \right]^2$ $\geq 0$. Adding these
two quantities, we get \eqref{temp15} and the result follows.
In particular, since $\log$ is an operator monotone function, the claim implies that 
\beq
\label{logpositivity}
\tr P\rho\left( \log\rho -\log\sigma\right) \geq 0.
\enq
\end{proof}

\begin{lemma}
\label{togetA}
Let $\rho, \sigma \geq 0$, $P = P_{\{\rho-\gamma \sigma\geq 0\}}$, and
$\log\gamma > S(\rho || \sigma)$, then
\beq
\label{temp5}
\tr P\rho \leq \frac{ \tr \rho \left( \log\rho-\log\sigma \right)^2
- \left[ S(\rho || \sigma) \right]^2 } {\left[\log\gamma - S(\rho || \sigma) \right]^2}.
\enq
\end{lemma}
\begin{proof}
We can rewrite \eqref{temp5} as
\beq
\tr P\rho \leq \frac{ \tr \rho \left\{ \log\rho-\log(\gamma \sigma) +
\left[\log \gamma - S(\rho || \sigma) \right] \mathbbm{1} \right\}^2 }
{\left[\log\gamma - S(\rho || \sigma) \right]^2}.
\enq
It suffices to show that
\begin{align}
\tr P\rho
& \stackrel{a}{\leq} \frac{ \tr P \rho \left\{ \log\rho-\log(\gamma \sigma) +
\left[\log \gamma - S(\rho || \sigma) \right] \mathbbm{1} \right\}^2 }
{\left[\log\gamma - S(\rho || \sigma) \right]^2} \\
& =
\frac{ \tr P \rho \left[ \log\rho-\log(\gamma \sigma) \right]^2 } 
{\left[\log\gamma - S(\rho || \sigma) \right]^2} 
+ \frac{ \tr P\rho \left[ \log\rho-\log(\gamma \sigma) \right] } 
{\left[\log\gamma - S(\rho || \sigma) \right]}
+ \tr P\rho,
\end{align}
where in $a$, the sufficient condition is due to multiplication by $P$.
The first term is non-negative and the second term is non-negative
using Lemma \ref{positivity} and the inequality follows.
\end{proof}

\begin{lemma}[Quantum Sibson identity]
\label{qsibson}
For any quantum state $\rho^{AB}$ in system $AB$ and $D_\lambda$ to
be the \renyi divergence of order $\lambda$, we have
\beq
D_{\lambda} (\rho^{AB}||\mathbbm{1}\otimes \sigma^{B}) =
D_{\lambda}(\sigma^* ||\sigma^{B}) + \frac{\lambda}{\lambda-1}\log \tr \left[\tr_{A}
\left(\rho^{AB}\right)^{\lambda}\right]^{\frac{1}{\lambda}},
\enq
\beq
\mbox{where } ~~~~ \sigma^{*} =
\frac{\left[\tr_{A}\left(\rho^{AB}\right)^\lambda\right]^{\frac{1}{\lambda}}}{\tr\left[\tr_{A}\left(\rho^{AB}\right)^\lambda\right]^{\frac{1}{\lambda}}}.
\enq
\end{lemma}
\begin{proof}
For the classical Sibson identity, see Ref. \cite{sibson-1969}. Note that
\begin{align}
D_{\lambda} (\rho^{AB}||\mathbbm{1}\otimes \sigma^{B}) & = \frac{1}{\lambda-1} \log \tr
\left(\rho^{AB}\right)^{\lambda} [ \mathbbm{1}\otimes (\sigma^{B})^{1-\lambda} ] \\
& = \frac{1}{\lambda-1}\log\tr \,
\tr_A\left(\rho^{AB}\right)^{\lambda}(\sigma^{B})^{1-\lambda} \\
& = \frac{1}{\lambda-1}\log\tr \left(\sigma^{*}\right)^{\lambda}(\sigma^{B})^{1-\lambda}
+ \frac{\lambda}{\lambda-1}\log \tr \left[\tr_{A}\left(\rho^{AB}\right)^{\lambda}\right]^{{\frac{1}{\lambda}}} \\
& =  D_{\lambda}(\sigma^{*}||\sigma_{B}) + \frac{\lambda}{\lambda-1}\log\tr \left[\tr_{A}\left(\rho^{AB}\right)^{\lambda}\right]^{{\frac{1}{\lambda}}}.
\end{align}
\end{proof}

\begin{lemma}
\label{coh-lemma}
If $ \cR > \max_{\rho^{AA^\prime}} I(A\rangle B)_{\sigma}$ , then 
\beq
\label{result}
\exists ~~ t \in [-1/2,0), ~~ \mbox{such that } \forall ~ s \in (t,0), ~~~
-s\cR+\min_{\rho^{AA^\prime}} E_{0}(s,\cN)_\rho > 0.
\enq
\end{lemma}
\begin{proof}
The proof follows the same argument as Lemma $3$ in Ref. \cite{ogawa-1999-converse}.
Let $g(s,\rho^{AA^\prime}) := -s\cR+ E_{0}(s,\cN)_\rho$ and suppose that
$\cR >  \max_{\rho^{AA^\prime}} I(A\rangle B)_{\sigma}$.
Note that $\forall$ $\rho^{AA^\prime}$ we have $g(0,\rho^{AA^\prime}) = 0$ and 
\beq
\label{result1}
g^\prime(0,\rho^{AA^\prime}) = -\cR + I(A\rangle B)_{\sigma} < 0.
\enq
Now suppose that \eqref{result} does not hold. Then 
\beq
\forall ~ t \in [-1/2,0), ~~ \exists ~ s \in (t,0), ~~~~~~ \mbox{such that  } \min_{\rho^{AA^\prime}}g(s,\rho^{AA^\prime}) \leq 0.
\enq
Hence, there exists a real sequence $\left\{s_{n}\right\}$ and a sequence $\left\{\rho_{n}^{AA^\prime}\right\}\subset \cS(\cH_{AA^\prime})$
such that
\beq
s_n \in \left(-\frac{1}{n+1},0\right)~~~ \mbox{and} ~~~ g(s_n,\rho_{n}^{AA^\prime})\leq 0.
\enq 
Now since $\cS(\cH_{AA^\prime})$ is a compact set (see Ref. \cite{pra-compact}), there exists a
subsequence of $\{\rho_{n}^{AA^\prime}\}$ that converges to some $\rho_{\infty}^{AA^\prime}$
as $n \rightarrow \infty$. Without loss of generality we can assume that 
$\rho_{n}^{AA^\prime} \rightarrow \rho_{\infty}^{AA^\prime}$. From the mean value theorem, it follows that 
\beq
\label{derivative}
\forall n,~~ \exists ~ r_n\in(s_n,0),~~ \mbox{ such that   } g^\prime(r_n,\rho_{n}^{AA^\prime}) = \frac{g(0,\rho_{n}^{AA^\prime})-g(s_n,\rho_{n}^{AA^\prime})}{0-s_n} \geq 0.
\enq
Since $g^\prime(s,\rho^{AA^\prime})$ is a continuous function of $(s,\rho^{AA^\prime})$,
\eqref{derivative} yields $g^\prime(0,\rho^{AA^\prime}_\infty) \geq 0$, which contradicts
\eqref{result1}.
\end{proof}

\begin{lemma}
\label{afinite}
Consider a cq-state $\rho^{AB} = \sum_x P_X(x) \ket{x} \bra{x}^A \otimes \sigma_x^B$,
where $P_X$ is a probability distribution and $\sigma_x^B \in \cS(\cH_B)$.
Then for all such cq-states,
\beq
\tr \rho^{AB} \left[ \log\rho^{AB}-\log(\rho^A \otimes \rho^B) \right]^2
\leq g(|AB|) + g(|B|),
\enq
where for any $d \in \mathbb{N}$, $g(1) = 0$,
\beq
g(d) := \left\{
\begin{array}{ll}
0.563, ~~ & d = 2 \\
\log^2d, ~~ & d \geq 3.
\end{array}
\right.
\enq
\end{lemma}
\begin{proof}
It is not difficult to see that for $\rho^A = \tr_B \rho^{AB}$, $\rho^B = \tr_A \rho^{AB}$,
\begin{align}
\left[ \log\rho^{AB}-\log(\rho^A \otimes \rho^B) \right]^2 & =
\log^2\rho^{AB} - \log^2\rho^A \otimes {\mathbbm{1}} + {\mathbbm{1}} \otimes
\log^2 \rho^B \\
& ~~~ - 2 \sum_x \log P_X(x) \, \ket{x} \bra{x}^A
\otimes \log \sigma_x^B \\
& ~~~ - \sum_x \ket{x} \bra{x}^A \otimes \left( \log \sigma_x^B \log \rho^B +
\log \rho^B \log \sigma_x^B \right).
\end{align}
Each term above with the negative sign contributes negatively when we take the
trace and hence, neglecting these terms gives us an upper bound.
We are left with
\beq
\tr \rho^{AB} \log^2\rho^{AB} + \tr  \rho^{B} \log^2\rho^{B}.
\enq
Using the arguments in Appendix E in Ref. \cite{polyanskiy-2010-coding}, it follows
that for a quantum state $\sigma$ of dimension $d$,
$\tr \sigma \log^2\sigma \leq g(d)$. QED.
\end{proof}

\begin{lemma}
\label{maxent}
Let $\ket{\psi}^{XY_1 Y_2}$ be a maximally entangled state, i.e.,
\beq
\ket{\psi}^{X Y_1 Y_2} = \frac{1}{\sqrt{|Y_1||Y_2|}}
\sum_{i_1=1}^{|Y_1|} \sum_{i_2=1}^{|Y_2|} \ket{i_1 i_2}^X \ket{i_1 i_2}^{Y_1 Y_2},
\enq
where $|X| \geq |Y_1| |Y_2|$, $\{\ket{i_1 i_2}^X\}$ and $\{\ket{i_1 i_2}^{Y_1 Y_2}\}$
are any orthonormal bases in $\cH_X$ and $\cH_{Y_1} \otimes \cH_{Y_2}$
respectively. Then, $|Y_2| \, \rho^{X Y_1}$ is a projector where
$\rho^{X Y_1} = \tr_{Y_2} \psi^{X Y_1 Y_2}$.
\end{lemma}
\begin{proof}
It is easy to see that
\begin{align}
\rho^{XY_1} &= \frac{1}{|Y_1||Y_2|} \sum_{i_1,i_1^\prime=1}^{|Y_1|}
\sum_{i_2,i_2^\prime=1}^{|Y_2|} \ket{i_1 i_2} \bra{i_1^\prime i_2^\prime}^X
\otimes \tr_{Y_2} \left(\ket{i_1 i_2} \bra{i_1^\prime i_2^\prime}^{Y_1Y_2} \right).
\end{align}
To prove the claim, it suffices to show that
$\left( |Y_2| \, \rho^{X Y_1} \right)^2 = |Y_2| \, \rho^{X Y_1}$ or
\beq
\label{temp7000}
\frac{1}{|Y_1|} \sum_{j_1=1}^{|Y_1|} \sum_{j_2=1}^{|Y_2|}
\tr_{Y_2} \left(\ket{i_1 i_2} \bra{j_1 j_2}^{Y_1Y_2} \right)
\tr_{Y_2} \left(\ket{j_1 j_2} \bra{j_1^\prime j_2^\prime}^{Y_1Y_2} \right)
= \tr_{Y_2} \left(\ket{i_1 i_2} \bra{j_1^\prime j_2^\prime}^{Y_1Y_2} \right).
\enq
Now consider the Schmidt decomposition of $\ket{i_1 i_2}^{Y_1Y_2}$, i.e.,
\beq
\label{schimidt}
\ket{i_1 i_2}^{Y_1Y_2} = \sum_{k}\sqrt{\alpha_{k i_1 i_2}}\ket{k i_1 i_2}^{Y_1}
\ket{k i_2 i_2}^{Y_2}.
\enq
Substituting in \eqref{temp7000} and simplifying, we have
\begin{align}
\mbox{LHS of \eqref{temp7000}} & = \frac{1}{|Y_1|} \sum_{j_1 j_2, k, l, l^\prime}
\sqrt{\alpha_{k i_1 i_2} \alpha_{l^\prime j_1^\prime j_2^\prime} }
\alpha_{l j_1 j_2}
\braket{l^\prime j_1^\prime j_2^\prime}{l j_1 j_2}^{Y_2} \braket{l j_1 j_2}{k i_1 i_2}^{Y_2}
\ket{k i_1 i_2} \bra{l^\prime j_1^\prime j_2^\prime}^{Y_1} \\
& = \sum_{k, l^\prime}
\sqrt{\alpha_{k i_1 i_2} \alpha_{l^\prime j_1^\prime j_2^\prime} } \,
\bra{l^\prime j_1^\prime j_2^\prime}^{Y_2}
\left( \frac{1}{|Y_1|} \sum_{j_1,j_2,l} \alpha_{l j_1 j_2}
\ket{l j_1 j_2} \bra{l j_1 j_2}^{Y_2} \right) \ket{k i_1 i_2}^{Y_2} \nonumber \\
& \hspace{4cm}  \ket{k i_1 i_2} \bra{l^\prime j_1^\prime j_2^\prime}^{Y_1} \\
& \stackrel{a}{=} \sum_{k, l^\prime}
\sqrt{\alpha_{k i_1 i_2} \alpha_{l^\prime j_1^\prime j_2^\prime} }
\bra{l^\prime j_1^\prime j_2^\prime}^{Y_2} \ket{k i_1 i_2}^{Y_2}
\ket{k i_1 i_2} \bra{l^\prime j_1^\prime j_2^\prime}^{Y_1} \\
& = \mbox{RHS of \eqref{temp7000}},
\end{align}
where in $a$, we have replaced the term inside the parenthesis by an identity
matrix since
\begin{align}
\frac{1}{|Y_1|} \sum_{j_1,j_2,l} \alpha_{l j_1 j_2}
\ket{l j_1 j_2} \bra{l j_1 j_2}^{Y_2} & = \frac{1}{|Y_1|} \sum_{j_1,j_2}
\tr_{Y_1} \left( \ket{l j_1 j_2} \bra{l j_1 j_2}^{Y_1Y_2} \right) = \mathbbm{1}.
\end{align}
QED.
\end{proof}

\section{Proof of Theorem \ref{qexp}}

Note that \eqref{zero} follows easily. We now verify \eqref{coherent} using the following differentiation rule (Lemma 4 in Ref. \cite{ogawa-1999-converse})
for a Hermitian operator $X(s)$ parametrized by a real parameter $s$
\beq
\label{rule}
\frac{\partial}{\partial s}\tr g[X(s)] = \tr g^\prime[X(s)] \frac{\partial X(s)}{\partial s}.
\enq
Let the spectral decomposition of $\sigma^{AB}$ be
$\sigma^{AB} = \sum_{i}\lambda_i\ket{i} \bra{i}^{AB}$
and let $\sigma_i = \tr_{A} \ket{i} \bra{i}^{AB}$.
Hence, we get
$\sigma^{B} = \tr_{A}\sigma^{AB} = \sum_{i}\lambda_{i}\sigma_i$ and
$\kappa_{1} := \tr_A(\sigma^{AB})^{1/(s+1)} = \sum_{i}\lambda_{i}^{1/(s+1)}
\sigma_i$.
It is easy to see using \eqref{rule} that
${\partial \kappa_1}/{\partial s} = -{\kappa_2}/{(s+1)}$,
where $\kappa_2 = \sum_{i}\lambda_i ^{\frac{1}{s+1}}\log (\lambda_i ^{\frac{1}{s+1}})
\sigma_i$. It now follows that 
\begin{align}
\frac{\partial g(s)}{\partial s} & =
\frac{\tr \kappa_{1}^{s}(\kappa_2-\kappa_1\log \kappa_1)}{\tr\kappa_1^{s+1}}, \\
\frac{\partial g(s)}{\partial s} \Big |_{s=0} &= \tr \Big[ \sum_{i}\lambda_{i} (\log \lambda_i)
\sigma_i- \big( \sum_i \lambda_i\sigma_i \big) \log \big(\sum_{i}\lambda_i\sigma_i \big)\Big],\\
& = H(B)_{\sigma}-H(A,B)_{\sigma},\\
& = I(A\rangle B)_{\sigma}.
\end{align}
Now we show that $g(s) + (s+1) \log |A|$ is an increasing function in $s$. Consider
the operators $E_i = \sqrt{\sigma_i/|A|}$. Then $\sum_i E_i^\dagger E_i =$
$\sum_i \tr_A \ket{i} \bra{i}^{AB}/|A|$ $= {\mathbbm{1}}$. Since $x^\gamma$,
$\gamma \in (0,1]$ is operator concave, we have, using the operator Jensen's
inequality and for $1/2 \leq \alpha \leq \beta < 1$, $\gamma = \alpha/\beta$,
\begin{align}
\left( \frac{1}{|A|} \sum_i \lambda_i^{1/\beta} \sigma_i \right)^\beta &
\leq \left( \frac{1}{|A|} \sum_i \lambda_i^{1/\alpha} \sigma_i \right)^\alpha,
\end{align}
or $g(\alpha-1) + \alpha \log |A| \leq g(\beta-1) + \beta \log |A|$.

\end{document}